\documentclass[twocolumn,showpacs,preprintnumbers,amsmath,amssymb,superscriptaddress]{revtex4}
\usepackage{dcolumn}
\usepackage{amsmath}
\usepackage{graphicx}
\usepackage{latexsym}
\usepackage{amsfonts}
\usepackage{amssymb}
\usepackage{color}
\usepackage{ulem}
\DeclareGraphicsExtensions{.pdf,.gif,.jpg}
\newcommand{\be}{\begin{equation}}
\newcommand{\ee}{\end{equation}}
\newcommand{\beq}{\begin{eqnarray}}
\newcommand{\eeq}{\end{eqnarray}}

\begin{document}

\newcommand\xmas{\mbox{$x_{\rm mas}$}}
\newcommand\rmas{\mbox{$r_{\rm mas}$}}
\newcommand\Xmas{\mbox{$X_{\rm mas}$}}
\newcommand\Hinf{\mbox{${\rm H}_{\rm inf}$}}
\newcommand\Omegainf{\mbox{$\Omega_{\rm inf}$}}
\newcommand\Omegabg{\mbox{$\Omega_{\rm bg}$}}
\newcommand\Hbg{\mbox{${\rm H}_{\rm bg}$}}
\newcommand\rhobg{\mbox{$\rho_{\rm bg}$}}
\newcommand\rhorbg{\mbox{$\rho^{\rm bg}_{\rm r}$}}
\newcommand\rhoinf{\mbox{$\rho_{\rm inf}$}}
\newcommand\kinf{\mbox{$k_{\rm inf}$}}
\newcommand\kbg{\mbox{$k_{\rm bg}$}}
\newcommand\ainf{\mbox{$a_{\rm inf}$}}
\newcommand\abg{\mbox{$a_{\rm bg}$}}
\newcommand\xpbg{\mbox{$x_{\rm p}^{\rm bg}$}}
\newcommand\xpinf{\mbox{$x_{\rm p}^{\rm inf}$}}
\newcommand\xmasbg{\mbox{$x_{\rm mas}^{\rm bg}$}}
\newcommand\xmasinf{\mbox{$x_{\rm mas}^{\rm inf}$}}
\newcommand\thetainf{\mbox{$\theta^{\rm inf}$}}
\newcommand\thetabg{\mbox{$\theta^{\rm bg}$}}
\newcommand\xHbg{\mbox{$1/{\rm H}_{\rm bg}$}}
\newcommand\xHinf{\mbox{$1/{\rm H}_{\rm inf}$}}

\title{Electric control of the magnetization in BiFeO$_{3}$/LaFeO$_{3}$ superlattices}

\author{Zeila Zanolli}
\affiliation{Physique Th\'eorique des Mat\'eriaux, Universit\'e de Li\`ege, B-4000 Sart Tilman, Belgium}

\author{Jacek C. Wojde\l} 
\author{Jorge \'{I}\~{n}iguez}
\affiliation{Institut de Ciencia de Materials de Barcelona (ICMAB--CSIC), Campus UAB, 08193 Bellaterra, Spain}

\author{Philippe Ghosez}
\affiliation{Physique Th\'eorique des Mat\'eriaux, Universit\'e de Li\`ege, B-4000 Sart Tilman, Belgium}
%\altaffiliation[Current address: ] {University of Liege (ULg), B-4000 Sart Tilman, Liege, Belgium.}

\date{\today}

\begin{abstract}
First-principles techniques are used to investigate the behavior of BiFeO$_{3}$/LaFeO$_{3}$ perovskite oxide superlattices epitaxially grown on a (001)-SrTiO$_3$ substrate. The calculations show that 1/1 superlattices exhibit a $Pmc2_1$ ground state combining a trilinear coupling of one polar and two oxygen rotational lattice modes, and weak ferromagnetism. The microscopic mechanism allowing one to manipulate the magnetization with an electric field in such systems is presented and its dependence on strain and chemical substitution is discussed. BiFeO$_{3}$/LaFeO$_{3}$ artificial superlattices appear to be good candidates to achieve electric switching of magnetization at room temperature. 
\end{abstract}

%\vspace{0.3cm}


\maketitle

%%% Introduction

During the last decade, the search for new magnetoelectric (ME) multiferroics capable of operating at room temperature has been the subject of intensive researches motivated, in particular,  by the dream of realizing magnetically readable and electrically writable data storage systems. In spite of continuous efforts, BiFeO$_3$ still remains to date the most promising room-temperature single-phase ME multiferroic. Recently, electric control of the magnetization has been demonstrated in BiFeO$_3$ \cite{ChuNature08} but, still, the identification of alternative systems suitable for practical applications remains a real challenge.

BiFeO$_3$ belongs to the family of multifunctional ABO$_3$ perovskite compounds that exhibit a wide variety of physical properties -- as ferroelectricity, piezoelectricity, (anti)ferromagnetism, colossal magnetoresistance, superconductivity, spin-dependent transport -- and have  already led  to numerous technological applications in electronics, optoelectronics, sensing, and data storage. Initially restricted to simple cubic perovskites and solid-solutions, the interest  of the researchers recently extended to artificially or naturally layered perovskite compounds, that can host even more exotic phenomena  \cite{Zubko_review11}.

In 2008, a new type of improper ferroelectricity, arising from an unusual trilinear coupling between one ferroelectric (FE) and two antiferrodistortive (AFD) motions, was reported in PbTiO$_3$/SrTiO$_3$ superlattices (SL) \cite{Bousquet_Nature08}. Since then, the coupling of lattice modes in perovskite layered structures has generated an increasing interest \cite{Ghosez_NatMat11}. The concept of ``hybrid improper ferroelectricity'' has been introduced \cite{Fennie_PRL11} and rationalized \cite{Stengel_PRB2012}, guiding rules to identify alternative hybrid improper ferroelectrics have been proposed \cite{Rondinelli_AdvMat2012}, and the emergence of ferroelectricity in rotation-driven ferroelectrics was discussed \cite{Mulder}.

It had been suggested that the trilinear coupling between FE and AFD structural degrees of freedom observed in PbTiO$_3$/SrTiO$_3$ SL might be a convenient way to tune the magneto-electric response in related magnetic systems \cite{Bousquet_Nature08}.  Subsequently, Benedek and FennieÊ~\cite{Fennie_PRL11} reported an unprecedented electric field control of the magnetization in Ca$_3$Mn$_2$O$_7$, a naturally-occurring Rudlesden-Popper layered compound in which a similar trilinear coupling of lattice modes can likely mediate an electric-field switching of the magnetization.  This mechanism has not yet been confirmed experimentally, but appears as one of the most promising ways to achieve strong magneto-electric coupling \cite{Birol_2012}. In Ca$_3$Mn$_2$O$_7$, however,  the magnetic order disappears at low temperature.

In this Letter, we investigate the behavior of a SL consisting of alternating monolayers of BiFeO$_3$ and LaFeO$_3$ (1/1 SL) epitaxially grown on (001)-SrTiO$_3$. 
Our choice is motivated by several arguments. First, BiFeO$_3$ and LaFeO$_3$  are G-type antiferromagnets (AFM) with comparable Fe magnetic moments  ($3.75 \mu_B$ \cite{Sosnowska_2002} and $4.6  \mu_B$ \cite{Koehler_1957}, respectively) and high N\'eel temperatures (643 K \cite{Kiselev_1963} and 750 K  \cite{Koehler_1957}, respectively). Hence, we can expect the SL to keep a similar magnetic order well above room temperature. Second, both materials have a strong tendency to $Pnma$-type distortion. Bulk LaFeO$_3$ crystallizes in the $Pnma$ phase at room temperature \cite{Geller_1956}. Bulk BiFeO$_3$ has a $Pnma$ phase \cite{Dieguez_PRB2011} very close in energy to the $R3c$ ground state \cite{Michel_1969}. This is in line with the design rule proposed by Rondinelli and Fennie \cite{Rondinelli_AdvMat2012} to achieve a $Pmc2_1$ ground-state in the SL which, in turn, is compatible with our targeted trilinear coupling of lattice modes.  A recent first-principles study of the Bi$_{1-x}$La$_{x}$FeO$_3$ solid solution further support that it might be the case \cite{Gonzalez_PRB2012}. We notice that, in both compounds, the distortions are strong and the structural phase transitions occur at high temperatures, a feature that could  likely be shared also by the SL.  
Finally, the polarization in such SL is expected to arise from the opposite motion of the sublattices of A and A' cations \cite{Mulder} and the distinct chemistry of Bi and La looks propitious to generate a sizable polarization. 
Here, we show that the BiFeO$_3$/LaFeO$_3$ 1/1 SL indeed combines a $Pmc2_1$ ferroelectric ground state with weak ferromagnetism and is a very promising candidate to achieve electric switching of polarization at room temperature.

%%%   Methods

Calculations were performed within density functional theory (DFT), using the ABINIT \cite{ABINIT} implementation of the projected augmented wave (PAW) method \cite{ABINIT_PAW}. 
We worked within the Local Spin Density Approximation (LSDA),  including a full rotationally invariant LSDA + U correction \cite{Liechtenstein_PRB95} with  $U = 4$ eV and $J = 0$ eV as Fe-atom on-site Coulomb and exchange parameters, respectively. All calculations were done considering a 40-atoms supercell which can be viewed as a $2\times2\times2$ repetition of the 5-atoms ABO$_3$ cubic perovskite primitive cell or, alternatively, as a $2\times2\times1$  repetition of the 10-atoms primitive cell of the $P 4/mmm$ parent structure of the 1/1 SL.
Convergence of the relevant physical properties was achieved for a kinetic energy cutoff of 18 Hartree and for a $2\times2\times2$ {\bf k}-point sampling of the Brillouin zone. Most calculations rely on a collinear spin approach and a G-type AFM ordering of Fe-atom spins. When specified, non-collinear spin calculations including spin-orbit coupling were performed using the VASP code \cite{VASP, VASP_PAW_PRB99} with $3\times3\times3$ {\bf k}-point mesh. Since the $J$ parameter can strongly affect the spin canting in these compounds \cite{Bousquet_PRB2010}, we kept the same $U-J = 4$ eV in these calculations but adopted the value $J = 0.5$ eV reproducing the experimental value of magnetization for BiFeO$_3$ ($0.05 \mu_B$ per Fe-atom \cite{Eerenstein_Science05}).

%%%  RESULTS

%1. Bulk BFO and LFO - to check the machinery

At first, we validated our approach by investigating the properties of G-type AFM  bulk  BiFeO$_3$ and LaFeO$_3$. We correctly reproduced the $R3c$ and  $Pnma$ ground state structures of BiFeO$_3$ and LaFeO$_3$, respectively. For BiFeO$_3$ a $Pnma$ phase is only 13 meV/f.u. above the ground state, consistently with Ref. \onlinecite{Dieguez_PRB2011}.  For LaFeO$_3$ a $R\bar{3}c$  structure (no polar distortion present) is 22 meV/f.u. above the ground state. The magnetic moment on the Fe atoms is $4.09 \mu_B$ in BiFeO$_3$ and $4.01  \mu_B$ in LaFeO$_3$, in agreement with experimental \cite{Sosnowska_2002, Koehler_1957} and previous first-principles LSDA + U  \cite{Neaton_PRB05, Yang_PRB1999} results. 

%2. BFO/LFO strained to STO (-1%). P4/mmm phonons 

Then, we modelled the BiFeO$_3$/LaFeO$_3$ 1/1 SL epitaxially grown on the [001] surface of a SrTiO$_3$  substrate. As a reference, we considered the high-symmetry paraelectric $P4/mmm$ structure. The mechanical constraint produced by the substrate was imposed by fixing the in-plane lattice constant of the SL ($a_0^{LDA} =$ 3.893 \AA) to the relaxed lattice constant of SrTiO$_3$  ($a_0^{LDA} =$ 3.847 \AA).  This corresponds to an epitaxial compressive strain of 1\% comparable to what is expected experimentally. After structural relaxation, the lattice parameter along the stacking direction was found to be $c_0^{LDA} =$ 7.762 \AA. This $P4/mmm$ phase is however not the lowest energy structure. Computation of the phonon modes at high-symmetry points of the Brillouin zone revealed the existence of many unstable modes, dominated by AFD motions (see \cite{Additional}).

\begin{figure}[htb]
\includegraphics[width=8.5 cm]{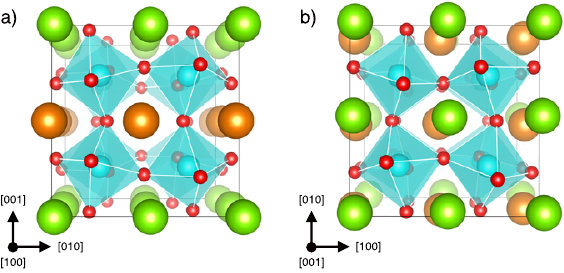}
\caption{(color online).  
$Pmc2_1$ structure of a BiFeO$_3$/LaFeO$_3$  1/1 SL epitaxially grown on (001)-SrTiO$_3$.
The O atoms (red) form corner-sharing octahedra, centered on Fe (cyan). Bi (green) and La 
(orange) layers alternate along the [001] direction. The two views evidence (a) anti-phase rotations 
of the oxygen octahedra about the [100] axis and (b) their in-phase rotations about the [001] 
axis. }
\label{fig:structure} 
\end{figure}

{\it Structural ground state --}  In order to identify the ground state structure we performed {\it ab initio} relaxation of various distorted structures arising from the condensation of different plausible combinations of unstable modes (see \cite{Additional}).  The lowest energy structure we found has a $Pmc2_1$ symmetry  and it is illustrated in Fig.~\ref{fig:structure}.  

Group theory analysis tells us that the atomic distortion $\Delta_{Pmc2_1}$ linking the $Pmc2_1$ ground state to the $P4/mmm$ reference phase can involve three distinct atomic distortions: two independent zone-boundary motions transforming respectively like the irreducible representations M$_5^-$ (2-dimensional) and M$_2^+$ (1-dimensional) and one zone-center motion transforming like the irreducible representation $\Gamma_5^-$ (2-dimensional). Although they also involve small cationic displacements, the M$_5^-$ and  M$_2^+$ distortions primarily correspond to oxygen motions. The M$_5^-$ distortion combines anti-phase tilts of the oxygen octahedra along the in-plane $x$ and $y$ directions ($a^-a^-c^0$ in Glazer's notation \cite{Glazer}) and will be labelled $\phi_{xy}^-=\phi_{x}^-+\phi_{y}^-$. The M$_2^+$ distortion corresponds to in-phase rotations of the oxygen octahedra along the $z$ direction ($a^0a^0c^+$) and will be labelled $\phi_{z}^+$. Finally, the polar $\Gamma_5^-$ distortion  induces a polarization along the $xy$ direction and will be labelled $P_{xy}=P_x+P_y$. Decomposition of $\Delta_{Pmc2_1}$ into these individual motions reveals that they all contribute significantly ($\phi_x^-, \phi_y^-, \phi_z^+, P_x, P_y \approx 20\%$). The $Pmc2_1$ phase appears therefore as the combination of a pattern of rotations and tilts of oxygen octahedra similar to the $Pnma$ phase of simple perovskites ($a^-a^-c^+$, with $a^-= 9.4^{\circ}$ and $c^+ = 9.5^{\circ}$) with a polar distortion, yielding a sizable polarization of 11.6 $\mu$C/cm$^2$  along the [110] direction (see \cite{Additional}). 

\begin{figure}[h]
%\begin{minipage}[htb]{1.0\textwidth}
\includegraphics[width=8.5 cm]{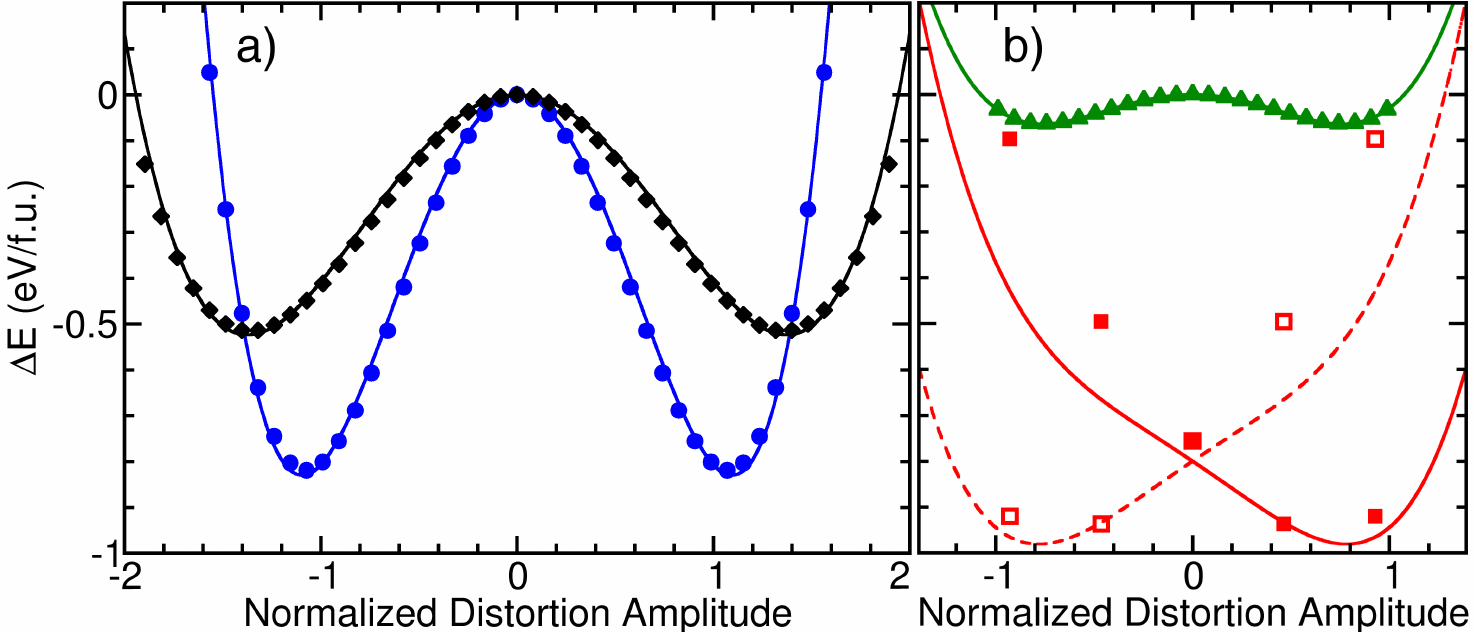}
\caption{Evolution of the energy (eV/f.u.) with the amplitude of 
the $\phi_{xy}^-$ (circles), $\phi_{z}^+$ (diamonds) and $P_{xy}$ (triangles) structural degrees of freedom. %order parameters. 
The $P4/mmm$ phase is taken as reference.
Symbols are first-principles results and lines our Landau-type model \cite{Additional}. Symmetric double wells are obtained for  individual distortions (in the absence of the others). The asymmetric well (filled squares) corresponds to the evolution of the energy with $P_{xy}$ when $\phi_{xy}^-$ and $\phi_{z}^+$ condense with the same amplitude as in the ground-state. The dashed curve (open squares) is for either $\phi_{xy}^-$ or $\phi_{z}^+$ reversed (opposite domain). 
}
\label{fig:improperFE} 
%\end{minipage}
\end{figure}

The structural distortions $\phi_{xy}^-$ and $\phi_{z}^+$ are essentially linked to the most unstable M$_5^-$ (241$i$ cm$^{-1}$) and M$_2^+$ (213$i$ cm$^{-1}$) zone-boundary phonon modes of the $P4/mmm$ phase.  The evolution of the energy with the amplitudes of  $\phi_{xy}^-$ and $\phi_{z}^+$ is illustrated in Fig. \ref{fig:improperFE}a, revealing double-well shapes able to produce substantial gains of energy. 

The weakly polar distortion $P_{xy}$ results from the combination of two strongly polar $\Gamma_5^-$ modes: one unstable (168$i$ cm$^{-1}$) corresponding to a Bi motion against  O, and another one stable (70 cm$^{-1}$), oppositely polarized, and dominated by a motion of La against O and Bi. Amazingly, in the combination of these two modes, oxygens' motion cancels almost exactly so that $P_{xy}$ corresponds to an opposite motion of Bi and La sublattices \cite{Mulder}. The evolution of the energy with the amplitude of $P_{xy}$  also corresponds to a double well (Fig. \ref{fig:improperFE}b, dots) but much weaker than with $\phi_{xy}^-$ and $\phi_{z}^+$.

%\begin{widetext}
%\begin{center}
%\end{center}
%\end{widetext}

A central point in the present study is the peculiar relationship between M$_5^-$, M$_2^+$ and $\Gamma_5^-$ irreducible representations of the $P4/mmm$ space group. The latter is compatible with the presence of a trilinear term in the expansion of the energy around the reference phase in terms of the $\phi_{xy}^-$, $\phi_{z}^+$ and $P_{xy}$ degrees of freedom:
\begin{equation}
E_{trilinear} =  - \lambda \phi_{xy}^-\phi_{z}^+P_{xy} 
\end{equation}

Linked to the presence of this term, it might be questioned if the system will behave like an hybrid improper ferroelectric \cite{Fennie_PRL11} in which the polarization is driven by the oxygens motions through the trilinear coupling.  From the previous discussion, the SL already exhibits weak proper FE instability in its $P4/mmm$ phase but, at the same time, the structural transition to the ground state is expected to be dominated by the stronger  $\phi_{xy}^-$ and $\phi_{z}^+$ instabilities. 
A  Landau-type model based on a fourth-order expansion of the energy in terms of $\phi_{xy}^-$, $\phi_{z}^+$ and $P_{xy}$ (see \cite{Additional}) reveals that the three degrees of freedom compete together (positive bi-quadratic couplings). The competition between $\phi_{xy}^-$ and $\phi_{z}^+$ seems however not sufficient to suppress completely any of them so that, together, they could bring the system in the $Pmc2_1$ symmetry and produce improper ferroelectricity as discussed in Ref. \onlinecite{Rondinelli_AdvMat2012}. We need however to take this result with caution since our model energy is restricted to a limited subspace, not including anharmonic couplings with other modes. Establishing unambiguously the improper ferroelectric character of the system would require a calculation under open-circuit electrical boundary conditions ($D=0$) as proposed in Ref. \cite{Stengel_PRB2012}. This is however beyond the scope of the present paper. 

Independently of the exact nature of the ferroelectric transition, the property of the SL that we would like to exploit  is  the trilinear coupling of lattice modes in the $Pmc2_1$ phase. Consistently with Eq. (1) and as illustrated in Fig. \ref{fig:improperFE}b, in this phase, the rotational modes $\phi_{xy}^-$ and $\phi_{z}^+$ act as a strong effective field $E_{eff} = \lambda \phi_{xy}^-\phi_{z}^+ =$ 6.72 MV/cm that distorts the $P_{xy}$ well in order to stabilize a given polarization. Our model energy qualitatively reproduces the first-principles data. However, it also points out that the inclusion of higher-order couplings is needed to quantitatively reproduce the energetics of the system. 
Importantly, in the $Pmc2_1$ ground state, the $P_{xy}$ well is highly asymmetric (red full line in Fig. \ref{fig:improperFE}b). Hence, switching the polarization mandatory requires the simultaneous switching of either $\phi_{xy}^-$ or $\phi_{z}^+$ (red dashed line in Fig. \ref{fig:improperFE}b), producing a very peculiar link between polar and rotational modes.          

%5. Magnetism on Fe-ions, spin canting and weak magnetization. fig:spin

{\it Magnetic ground state --} At the magnetic level, both BiFeO$_3$ and LaFeO$_3$ exhibit a G-type AFM arrangement.  Consistently with that, our collinear spin calculations for the BiFeO$_3$/LaFeO$_3$ SL predict a G-type AFM order with magnetic moment on Fe-ions of 4.01 $\mu_B$. In the $Pmc2_1$ structure of the 1/1 SL, similarly to what happens in the $Pnma$ phase of single perovskites \cite{Bousquet_2011}, the spins are nevertheless not forced by symmetry to stay exactly anti-parallel but can rotate to produce a weak magnetic moment. To quantify this effect we included spin-orbit interaction and performed non-collinear spin calculations.

\begin{figure}[t]
%\begin{figure*}[th]
%\centering
\includegraphics[width=8.5 cm]{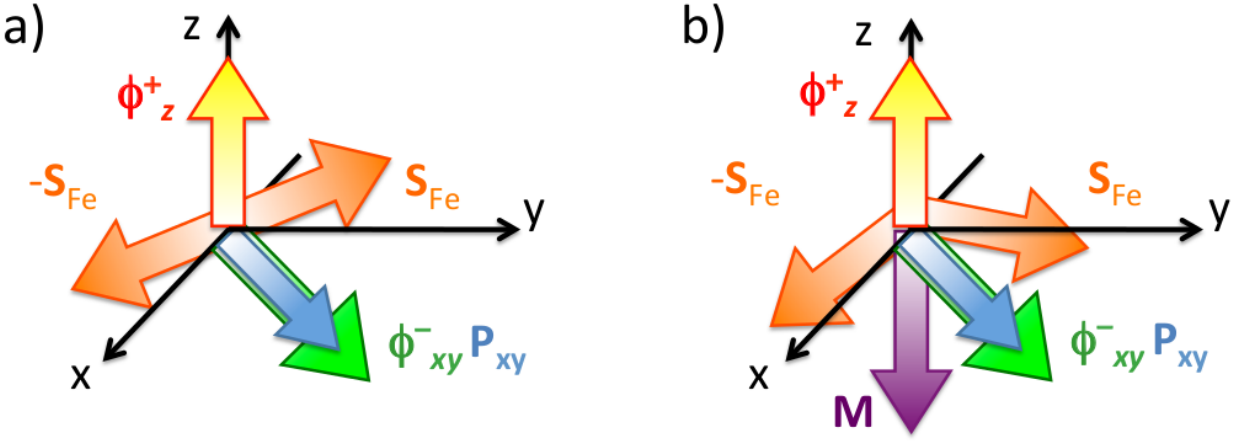}
\caption{Schematic view of the structural ($\phi_{xy}^-$, $\phi_{z}^+$, and  $P_{xy}$)  and magnetic (${\bf S}_{Fe}$, ${\bf M}$) degrees of freedom in the $Pmc2_1$ phase of the BiFeO$_3$/LaFeO$_3$ 1/1 SL. The easy axis for the spins ${\bf S}_{Fe}$ is in-plane and perpendicular to {\bf P}  (a).  Spin canting produces a weak magnetization ${\bf M}$ pointing in the out-of-plane direction (b).}
\label{fig:spin} 
%\end{figure*}
\end{figure}

These calculations revealed that the easy axis for the spins is along the [1 -1 0] direction, which is  in-plane and perpendicular to the polarization direction. Moreover, letting the spin relax gives rise to a net spin-canted moment (weak magnetization) $M = 0.43 \mu_B$ pointing in the [00-1] direction, orthogonal to both the polarization and the easy axis, as shown in Fig.~\ref{fig:spin}.

From knowledge of the magnetic space group, group theory analysis \cite{Bertaut} shows that the anti-phase rotations will control the sign of the canted moment in a material like ours.
In order to concretely attest for that, we repeated the previous calculations for different ferroelectic domain configurations, switching $P_{xy}$ and either $\phi_{xy}^-$ or $\phi_{z}^+$.  We found that switching $P_{xy}$ together with $\phi_{xy}^-$ switches the direction of the weak magnetization from [00-1] to [001]. Switching $P_{xy}$ together with $\phi_{z}^+$ leaves  ${\bf M}$ unchanged, instead. This  demonstrates that electric switching of the magnetization is {\it a priori} achievable in this type of system, as long as the switching of the polarization is accompanied by the switching of  $\phi_{xy}^-$. In this context, clarifying how the system will switch in an electric field appears therefore an important issue.

%6. switching path search. fig: Ebarrier

{\it Switching path --} Ferroelectric switching is a complex dynamical effect, also strongly influenced by defects, and is certainly out of  the scope of first-principles calculations at 0 K. Nevertheless, adopting an approach that is similar to that of Ref.~\cite{Fennie_PRL11}, we can get some insight into the most likely intrinsic switching scenario. 

\begin{figure}[t]
%\begin{figure*}[th]
%\centering
\includegraphics[width=8.5 cm]{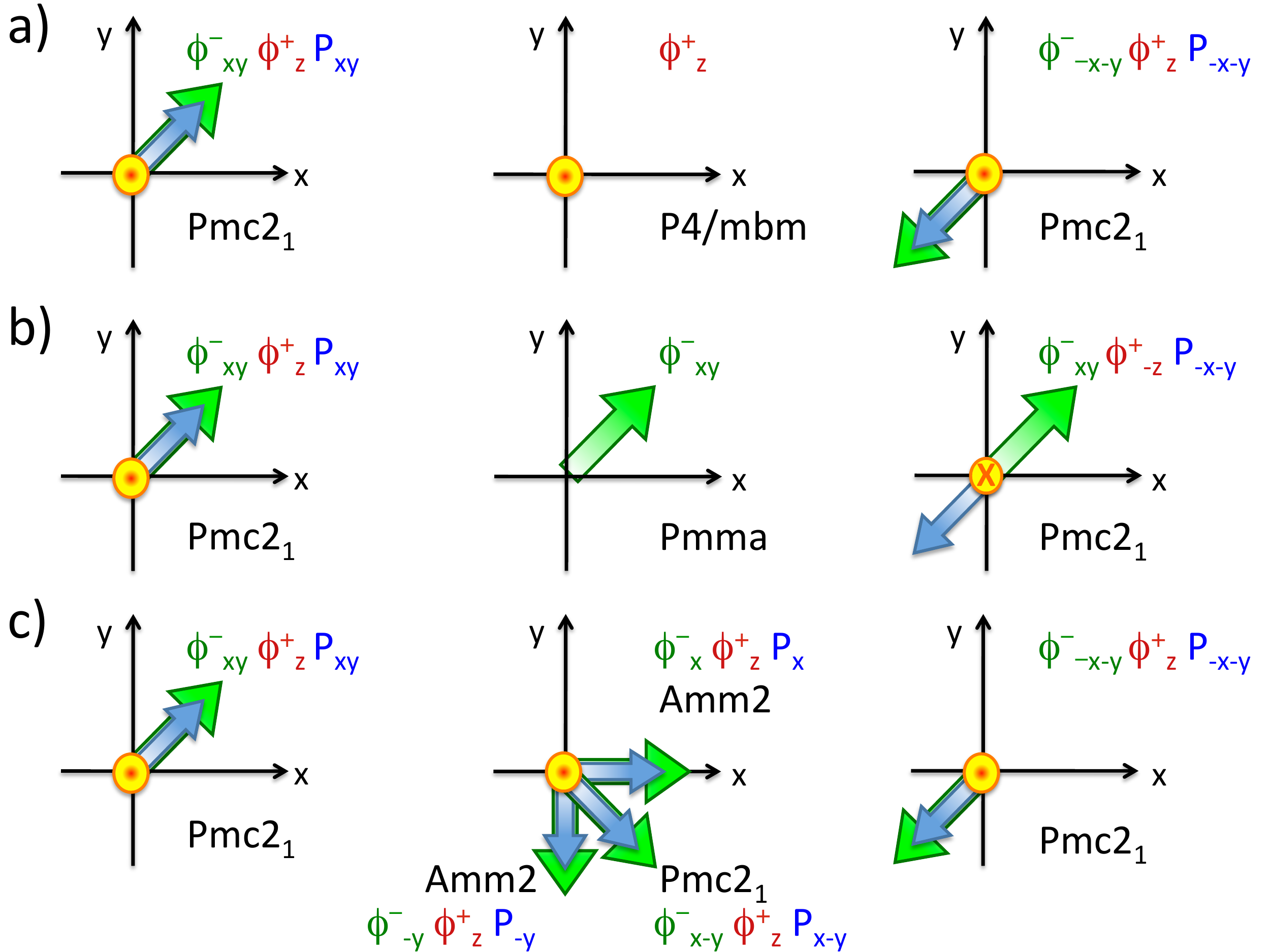}
\caption{Possible intrinsic ferroelectric switching paths in the $Pmc2_1$ phase of the BiFeO$_3$/LaFeO$_3$ 1/1 SL, labeled according to the name of intermediate high-symmetry phases: (a) $P4/mbm$, (b) $Pmma$, (c) $Amm2$.}
\label{fig:paths} 
%\end{figure*}
\end{figure}

\begin{figure}[t]
%\begin{figure*}[th]
%\centering
\includegraphics[width=7.5 cm]{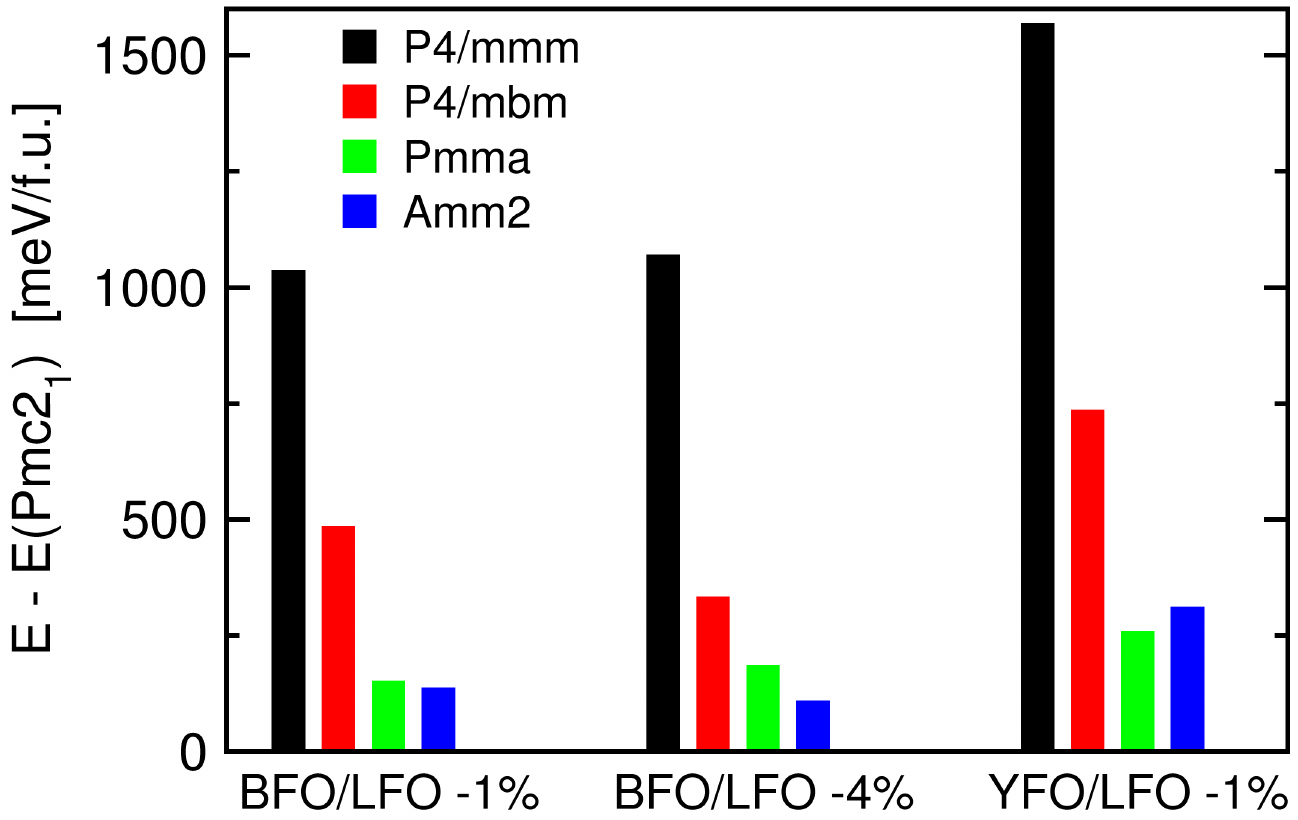}
\caption{Intrinsic energy barriers (meV/f.u.) associated to various ferroelectric switching paths in the $Pmc2_1$ phase of BiFeO$_3$/LaFeO$_3$ ($-1\%$ and $-4\%$ strain), and YFeO$_3$/LaFeO$_3$  ($-~1\%$ strain) 1/1 SL. The switching paths are labeled as in Fig.~\ref{fig:paths}.  The energy difference with respect to the parent structure $P4/mmm$ is reported for comparison.}
\label{fig:Ebarrier} 
%\end{figure*}
\end{figure}

More specifically, we can first restrict our investigations to the limited subspace spanned by $\phi_{xy}^-$, $\phi_{z}^+$, and  $P_{xy}$,  and compare the intrinsic energy barriers $\Delta E_i = E_i - E_{Pmc2_1}$ associated to the switching of {\bf P} through the intermediate $P4/mbm$ and $Pmma$ phases as illustrated in Fig.~\ref{fig:paths}.a-b. The energy barriers estimated from a full structural relaxation of the intermediate phases are $\Delta E_{P4/mbm}= 484$ meV/f.u. and $\Delta E_{Pmma}= 151$ meV/f.u., as summarized in Fig.~\ref{fig:Ebarrier}. These results are in line with the shallower well associated to the $\phi_{z}^+$ distortion in Fig. \ref{fig:improperFE}.a and suggest that the switching of {\bf P} might preferably be accompanied with that of $\phi_{z}^+$, which would leave {\bf M} invariant.   

In practice, the switching of {\bf P} via $P4/mbm$ or $Pmma$ phases with $P=0$ seems very unlikely, however. 
Remembering that $P_{xy}$ and $\phi_{xy}^-$ are 2-dimensional degrees of freedom, the rotation of {\bf P} within the $xy$-plane and its switching via an $Amm2$ phase, as illustrated in Fig.~\ref{fig:paths}.c,  appears to be a more realistic alternative. Indeed, we found that it is associated to a lower energy barrier $\Delta E_{Amm2}= 136$ meV/f.u.. Along this path $\phi_{xy}^-$ is reversed with $P_{xy}$, simultaneously allowing the targeted switching of {\bf M}.   

We repeated these calculations imposing an epitaxial compressive strain of -4\%. We found that increasing the strain lowers the energy barrier $\Delta E_{Amm2}$ to 108 meV/f.u., favoring this path even more with respect to the alternative investigated ones (Fig.~\ref{fig:Ebarrier}). 
We finally also checked  the effect of substituting BiFeO$_3$ with YFeO$_3$, a material having a $Pnma$ ground state and so {\it a priori} more in line with the energetic design rule outlined in Ref.~\cite{Rondinelli_AdvMat2012} to achieve hybrid improper ferroelectricity. We found, however, that in the YFeO$_3$/LaFeO$_3$ SL with -1\% strain  the energy barriers are significantly higher: $\Delta E_{Amm2}$ raises to 310 meV/f.u., now overcoming $\Delta E_{Pmma} = 258$ meV/f.u.. This suggests that, although enhancing the $Pnma$ character of the parent compounds might favor hybrid improper ferroelectricity, it is not necessarily useful to achieve electric switching of the magnetization.

We need to be careful in the interpretation of the previous results. We restricted our comparison to a limited number of ferroelectric switching paths. Also, the rotation of the polarization is far from being a rigid one as na\"ively suggested in Fig.~\ref{fig:paths}.c: in the relaxed $Amm2$ phase, the atomic distortions associated to $P_x$ and $\phi_x^-$ are substantially different than in the $Pmc2_1$ phase. This highlights the very complicated character of the energy landscape and the need of more complex simulations including all structural degrees of freedom to properly tackle this issue. 
Then, we make the assumption, implicit in previous works \cite{Fennie_PRL11,Birol_2012}, that the polarization switching occurs too rapidly for the spin structure to react to it: we assume that the spin structure after switching will closely resemble the starting configuration, except for the small canted component. Although this seems reasonable and such a final magnetic configuration corresponds to a minimum of the energy, this assumption cannot be validated from our static calculations.
Nevertheless, our investigations clarifies that (i) relatively low-energy intrinsic switching paths, comparable in magnitude with those of conventional ferroelectric like PbTiO$_3$, exist in such SL and (ii) these are compatible with the switching of the magnetization. 

In conclusion, we have shown that a BiFeO$_3$/LaFeO$_3$ 1/1 SL epitaxially grown on (001)-SrTiO$_3$ exhibits a multiferroic $Pmc2_1$ ground state combining ferroelectricity and weak ferromagnetism. In view of the high critical temperatures associated to the structural and magnetic phase transitions in the BiFeO$_3$ and LaFeO$_3$ parent compounds, we can expect that the SL will adopt this ground state well above room temperature. Our calculations highlighted the existence of a low energy ferroelectric switching path, along which the magnetization is also likely reversed. Although we cannot exclude alternative switching mechanisms, the presented results suggest that such system is a promising candidate to achieve electric switching of the magnetization at room temperature. We hope that this work will motivate further experimental investigations.

%==============      ACKNOWLEDGMENTS    =================
%\begin{acknowledgments}
All authors acknowledge financial support from European FP7 project OxIDes (Grant No. NMP3-SL-2009-228989). Work at ULG was also supported by the Concerted Research Action (ARC) TheMoTherm. Work at ICMAB was also funded by MINECO-Spain (Grants No. MAT2010-18113 and No. CSD2007-00041) and by CSIC (JAE-doc program).  Ph.G. acknowledges a Research Professorship from the Francqui Foundation (Belgium).

%\end{acknowledgments}

\end{document}